\documentclass[12pt,preprint]{aastex}
\usepackage{amsmath,amssymb,color}

\begin{document}
\title{Long Duration X-Ray Flash and  X-Ray Rich Gamma Ray Burst from Low Mass Population III Star}

\author{Daisuke Nakauchi$^{1}$, Yudai Suwa$^{2}$, Takanori Sakamoto$^{3,4,5}$, 
Kazumi Kashiyama$^{1,6}$, and Takashi Nakamura$^{1}$}

\affil{
$^{1}$Department of Physics, Kyoto University, Oiwake-cho, Kitashirakawa, Sakyo-ku,
Kyoto 606-8502, Japan\\
$^{2}$Yukawa Institute for Theoretical Physics, Kyoto University, Oiwake-cho, Kitashirakawa, Sakyo-ku, 
Kyoto 606-8502, Japan\\
$^{3}$Center for Research and Exploration in Space Science and Technology (CRESST), NASA Goddard Space Flight Center, Greenbelt, MD 20771\\
$^{4}$Joint Center for Astrophysics, University of Maryland, Baltimore County, 1000 Hilltop Circle, Baltimore, MD 21250\\
$^{5}$NASA Goddard Space Flight Center, Greenbelt, MD 20771\\
$^{6}$Department of Astronomy and Astrophysics, Pennsylvania State University, University Park, PA 16802, USA
}

\begin{abstract}
Recent numerical simulations suggest that Population III (Pop III)
stars were born with masses not larger than $\sim 100M_\odot$ but
typically $\sim 40M_{\odot}$. By self-consistently considering the jet generation and propagation in the envelope of
these low mass Pop III stars, we find that a Pop III blue super giant
star has the possibility to raise a gamma-ray burst (GRB) even though
it keeps a massive hydrogen envelope. We evaluate observational
characters of Pop III GRBs and predict that Pop III GRBs have the
duration of $\sim 10^5$ sec in the observer frame and the peak
luminosity of $\sim 5 \times 10^{50}\ {\rm erg}\ {\rm
  sec}^{-1}$. Assuming that the $E_p-L_p$ (or $E_p-E_{\gamma, \rm
  iso}$) correlation holds for Pop III GRBs, we find that the spectrum
peak energy falls $\sim$ a few keV (or $\sim 100$ keV) in the observer
frame. We discuss the detectability of Pop III GRBs by future
satellite missions such as {\it EXIST} and {\it Lobster}. If the
$E_p-E_{\gamma, \rm iso}$ correlation holds, we have the possibility
to detect Pop III GRBs at $z \sim 9$ as long duration X-ray rich GRBs
by {\it EXIST}. On the other hand, if the $E_p-L_p$ correlation holds,
we have the possibility to detect Pop III GRBs up to $z \sim 19$ as
long duration X-ray flashes by {\it Lobster}.
\end{abstract}
\maketitle

\section{Introduction}

Gamma ray bursts (GRB) are the brightest phenomena in the
universe. Long-soft type GRBs are considered to originate from deaths
of massive stars such as Wolf-Rayet (WR) stars
\citep{2011arXiv1104.2274H}. The most widely accepted scenario for
long GRBs is the collapsar scenario \citep{1993ApJ...405..273W,
  1999ApJ...524..262M}. In this model, after the gravitational
collapse of a massive stellar core, a black hole and an accretion disk
system is formed and it launches a relativistic jet by magnetic field
or neutrino pair annihilation process. If the jet can break out the
stellar envelope successfully, a GRB is raised by converting the jet
kinetic energy into the radiation energy.

Owing to their brightness and detections at high redshift universe,
GRBs are expected to be one of the powerful tools to probe the early
universe. The development of observational instruments and early
follow-up systems enable us to discover some high redshift GRBs. The
most distant one ever is GRB 090429B at $z=9.4$
\citep{2011ApJ...736....7C} and GRB 090423 at $z=8.3$
(e.g. \citealt{2009Natur.461.1254T, 2009Natur.461.1258S,
  2010ApJ...712L..31C}) follows it. If GRBs can be raised by first
stars and detected, we will draw informations about the early
universe, e.g., the star formation history and the reionization
history.

First stars in the universe, so called Population III~(Pop III) stars,
are considered to be formed from metal free gas in the very early
universe.  The metal-free primordial gas cools less efficiently
compared to the metal-contained present-day gas, which allows the
primordial gas to have larger fragmentation masses.  Since it was
considered that the whole fragmented gas clump collapsed to form a
single star, Pop III stars were theoretically predicted to be very
massive $\sim 100 - 1000 M_\odot$ \citep{2002Sci...295...93A,
  2002ApJ...564...23B}. However, recent studies suggest that this is
not always the case and that a massive gas clump can experience
further fragmentation to form a binary system
\citep{2009Sci...325..601T, 2010MNRAS.403..45,
  2011ApJ...727..110C}. \cite{2004ApJ...603..383T} and
\cite{2008ApJ...681..771M} suggested that the UV radiation from the
central protostar can ionize the surrounding neutral gas and suppress
the accretion onto the protostar. They analytically investigated this
feed back effect on the protostar evolution and found that Pop III
stars finally obtain mass typically $\sim 140 M_{\odot}$.  More
recently, \cite{2011Sci...334.1250H} performed two-dimensional
simulations of the protostar evolution including the above feed back
effect and found that Pop III stars finally obtain masses typically
$\sim 40 M_{\odot}$.  They also concluded that the UV radiation from
the central star eventually stops the mass accretion and the growth of
the star by the evaporation of the surrounding gas.

It is considered that metal free Pop III stars do not lose mass,
keeping large hydrogen envelopes until the pre-supernova stage,
because of the low opacity envelopes \citep{2002RvMP...74.1015W}.  The
final fate of a Pop III star depends on the stellar mass
\citep{2003ApJ...591..288H}. After the stellar core collapse, those
Pop III stars in the range of $10M_\odot \lesssim M \lesssim 25
M_\odot$ would explode as supernovae and form neutron stars as
remnants. Those in $25M_\odot \lesssim M \lesssim 40 M_\odot$ form
black holes as remnants after the fall back accretion of the envelopes
onto the temporally formed neutron stars. More massive stars
($40M_\odot \lesssim M \lesssim 140 M_\odot$ and $260M_\odot \lesssim
M$) would fail to blow out their envelopes and promtoly form massive
black holes, except those stars in $140M_\odot \lesssim M \lesssim 260
M_\odot$ who end as pair-instability supernovae due to the explosive
nucleosynthesis. These remnant massive black holes are expected to
raise various violent phenomena
\citep{2001ApJ...550..372F,2007PASJ...59..771S}.

There have been some studies about the productivity of GRBs from
massive Pop III stars ($\gtrsim 100M_{\odot}$)
\citep{2010ApJ...715..967M,2010MNRAS.402L..25K,2011ApJ...726..107S,2012ApJ...754...85N}.
The former two studies assumed a massive black hole surrounded by an
accretion disk as an outcome of a massive stellar collapse, and
estimated the accretion rate onto the black hole. Then they evaluated
the jet luminosity and showed that the burst activity of a Pop III
star is observable by current detectors.  In
\cite{2011ApJ...726..107S}, they analytically studied the jet
propagation in the stellar envelope and showed that massive Pop III
stars ($\sim 900 M_\odot$) can produce GRBs although they have large
hydrogen envelopes, since the long lasting accretion provides enough
energy and time for the successful jet breakout.  In addition,
\cite{2012ApJ...754...85N} performed two-dimensional relativistic
hydrodynamic simulations in which the accretion onto a black hole and
the jet production are treated in a self-consistent way for stellar
models of massive Pop III stars ($915M_\odot$), Wolf-Rayet stars
(initially $16 M_{\odot}$), and low mass Pop III stars
($40M_\odot$). They confirmed the validity of the analytic results in
\cite{2011ApJ...726..107S} and also found that $40 M_{\odot}$ Pop III
stars can be progenitors of GRBs, but did not study their
observational characteristics and detectability.

The idea of GRBs from blue super giants (BSG) was suggested in
\cite{2001ApJ...556L..37M}. Although they considered the jet dynamics
in the stellar envelope, they treated a steady jet and did not reflect
the central engine activity caused by the change of the accretion
rate. They did not evaluate the possibility of GRBs from BSGs
quantitatively. On the other hand, \cite{2012ApJ...752...32W}
discussed gamma-ray transients from Pop III BSG collapsars by
investigating the mass accretion of the outermost layers of a star,
but did not discuss the jet propagation and the jet break
out. Assuming that the conversion efficiency of the accretion energy
to the radiation energy $\sim 10^{-2}$, they found that Pop III BSGs
can produce long gamma-ray transients with duration $10^{4-5}$ sec and
luminosity $10^{48-49}$ erg ${\rm sec}^{-1}$.  In this paper, we
simultaneously investigate both aspects (the jet propagation and the
central engine activity) in a self-consistent way by including the
following physical processes; the stellar collapse, the non-steady
jet injection, and the jet propagation in the stellar envelope.  By
doing this, we quantitatively discuss the possibility of the jet break
out and GRB especially for low mass Pop III stars (around $40
M_{\odot}$).

In \S2, after introducing the stellar models and the jet propagation
models, we investigate the productivity of a GRB focusing on a $40
M_{\odot}$ Pop III star, which is a Pop III star with the typical mass
reported by the state-of-the-art simulation done in
\cite{2011Sci...334.1250H}. In \S3, we calculate the observational
characters, such as the duration $T_{90}$, the peak luminosity $L_p$
and the spectrum peak energy in the observer frame $E_p^{\rm obs}$, of
GRBs from $40 M_{\odot}$ Pop III progenitors. Then we evaluate the
detectability of such Pop III GRBs by future detectors such as {\it
  Lobster} and {\it EXIST} in detail, varying the redshift of a
burst. We apply the above discussions to different progenitor models
with masses of $30-90 M_{\odot}$. In the last part of \S3, we evaluate
the light curves of Pop III GRB radio afterglow emissions and their
detectability by the Low Frequency Array (LOFAR) and the Expanded Very
Large Array (EVLA).  \S4 is devoted to the summary and discussions.

\section{GRBs from low mass Pop III stars}
\subsection{Progenitor and Relativistic Jet models}
We employ a pre-collapse stellar model of z40.0 by
\cite{2002RvMP...74.1015W}, which provides the structure of a $40 M_{\odot}$ with zero metallicity at the final phase of the stellar evolution.
It is considered that a $40 M_\odot$ Pop III star promptly forms a
black hole after the core collapse \citep{2003ApJ...591..288H}. Then
we consider the subsequent evolution of the stellar collapse and the
jet propagation in the stellar envelope following the similar
prescription to \cite{2011ApJ...726..107S}.

We first assume that the collapse proceeds in a spherically symmetric
manner without pressure support so that each mass shell of the star
within mass [$M_r, M_r + dM_r$] and radius [$r, r+dr$] falls into the
central core in the free-fall time scale $t_{\rm ff} (r)=
\sqrt{r^3/GM_r}$. We calculate the mass accretion rate from the
expression $\dot{M} (r)= dM_r/dt_{\rm ff}(r)$.  When the mass of the
central core becomes $3M_\odot$, we identify that a black hole is
formed since the maximum possible mass of a neutron star is $\sim
3M_\odot$ \citep{1974PhRvL..32..324R, 1976ApJ...207..592C}.  After the
formation of the black hole, we assume that a cold relativistic jet
with a constant opening angle $\theta_j =5^{\circ}$ has been launched
and we take this moment as the origin of time ($t=0$).

There are mainly two candidates for the jet production mechanisms,
i.e., the neutrino annihilation process and the magnetic process. In
\cite{2011ApJ...726..107S}, they showed that the jet model based on
the neutrino annihilation process is not appropriate for producing
GRBs from very massive Pop III stars. Accordingly we adopt the jet
model based on the magnetic process. In this model, the jet injection
luminosity $L_{\rm jet}(t)$ is considered to be represented as $L_{\rm
  jet}(t) = \eta \dot{M}(t) c^2$ (see \citealt{2011ApJ...726..107S}
and references there in), where the constant $\eta$ is an energy
conversion efficiency and we take the value of $\eta = 6.2 \times
10^{-4}$. This is a calibrated value so as for Wolf-Rayet stars to
reproduce the energetics of canonical local long GRBs, i.e., nearly
$10^{52}$ ergs of energy should be injected into the relativistic jet
after the breakout.

In this paper it looks like that we neglect the effect of the
stellar rotation and treat the stellar collapse in spherically
symmetric way. According to \cite{2008MNRAS.388.1729K}, when the
stellar rotation is taken into consideration, the accretion time scale
onto the central BH for each mass shell $t_{\rm acc}(r)$ can be
represented as $t_{\rm acc} (r) \sim t_{\rm ff}(r)/\alpha$, where
$\alpha \sim 0.1$ is the standard dimensionless viscosity parameter of
the disk. As described in \cite{2011ApJ...726..107S}, we regard that
this uncertain factor is absorbed within the calibrated parameter
$\eta$. Therefore, we think that the disk formation is implicitly
taken into account and that the spherically symmetric prescription
makes sense.

\begin{figure}[!htb]
\includegraphics[angle=-90,scale=1.0]{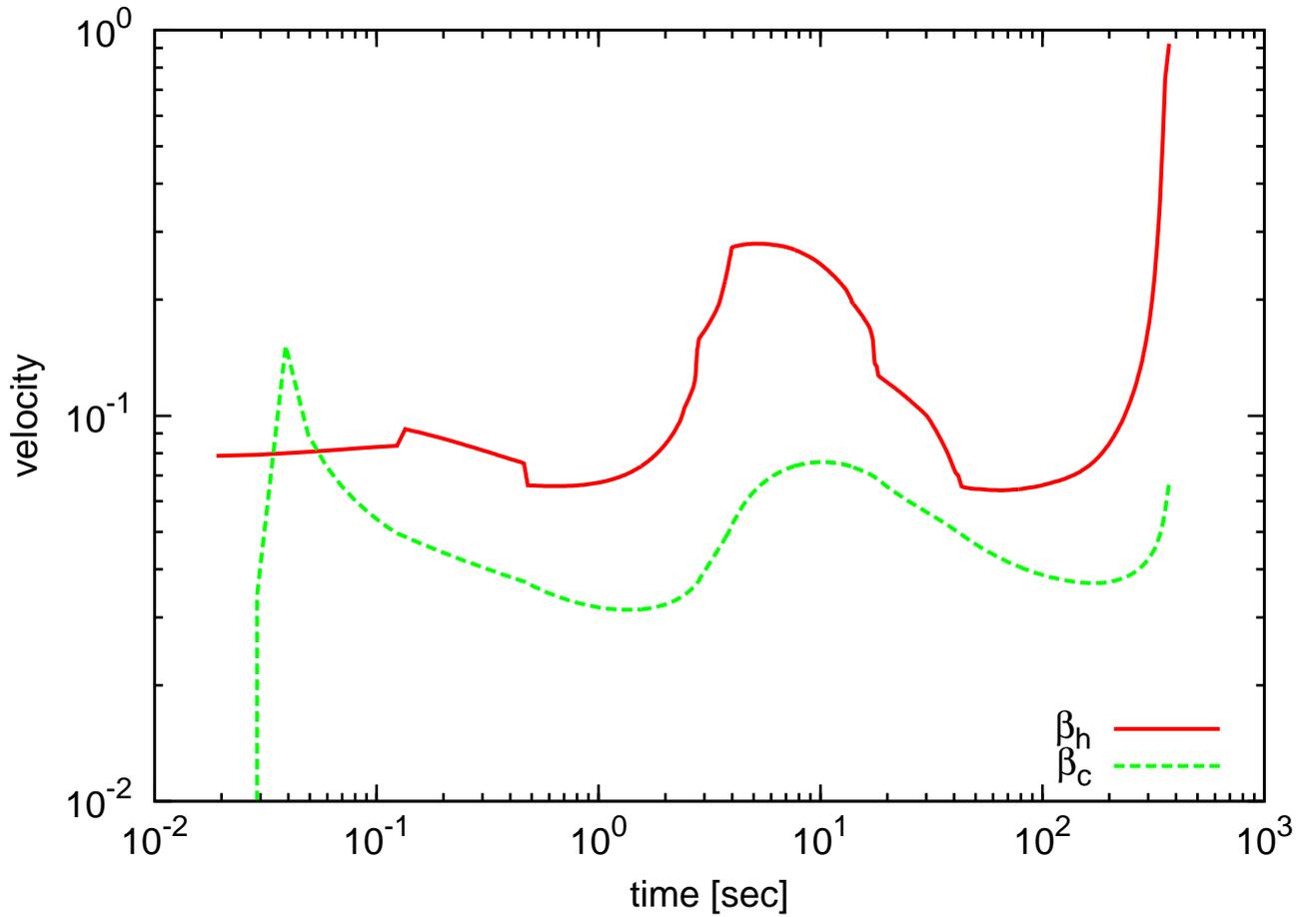}
\caption{The time evolution of the jet head velocity $\beta_h$ (the
  red solid line) and the cocoon velocity $\beta_c$ (the green dashed
  line) during the propagation in the stellar envelope. The abscissa
  is the time from which the jet is activated. The progenitor is a
  $40M_{\odot}$ Pop III star.}
\label{fig:vel}
\end{figure}

In the following sections, we consider the propagation of a jet in the
stationary stellar envelope. For the $40 M_{\odot}$ Pop III model, the
He core mass $M_{\rm He}$ and the He core radius $r_{\rm He}$ are
$M_{\rm He} \sim 22 M_{\odot}$ and $r_{\rm He} \sim 10^{11}$ cm,
respectively.  Therefore, the collapse time scale of the He core is
estimated as $t_{\rm coll} \sim 600$ sec. On the other hand, the time
scale for the jet head to reach the outer edge of the He core can be
evaluated as $t_{\rm cross} \sim r_{\rm He} / 0.1 c \sim 30$ sec,
since we can see that the average velocity of the jet head within the
He core is $\sim 0.1 c$ (see Fig. \ref{fig:vel}). Accordingly, $t_{\rm
  cross} \ll t_{\rm coll}$ holds. We confirm that this inequality
holds better in outer layers and that calculations based on the
stationary envelope is self consistent.

\subsection{Jet Propagation in the Pop III Star Envelope}
First, we consider the propagation of a jet in the stellar
envelope. As pointed out in the previous subsection, we approximate
that the stellar envelope is stationary and that the density profile
is the same as that in the pre-supernova stage until the jet break out.  A
jet propagating through the stellar envelope forms forward and reverse
shocks at its head. Here, we assume that the separation between these
shocks is small compared to the distance from the stellar center.
From the continuity of the momentum flux at the jet head, we have
\citep{2003MNRAS.345..575M},
\begin{equation}
\rho_j c^2 h_{j} (\Gamma_{j} \Gamma_{h})^2(\beta_j-\beta_h)^2 + P_j = \rho_{\ast} c^2 h_{\ast} (\Gamma_{h}\beta_h)^2 + P_{\ast} ,
\label{eq:pressure balance}
\end{equation}
where $\rho, h, \Gamma, \beta $ and $P$ represent the density, the
specific enthalpy, Lorentz factor, the velocity divided by the speed
of light $c$, and the pressure, respectively.  The subscripts j, h and
$\ast$ stand for the jet, the jet head and the stellar envelope,
respectively.  We consider a cold jet so that we can neglect $P_j$ in
l.h.s. of Eq. \eqref{eq:pressure balance}. In r.h.s. of
Eq. \eqref{eq:pressure balance}, we can approximate $h_{\ast} \sim 1$
and neglect $P_{\ast}$, since stellar material is
non-relativistic. Then the velocity of the jet head ($\beta_h$) is
expressed as
\begin{equation}
\beta_h(t) =  \beta_j \left[1+ \left[ \frac{\pi {r_h}^2 {\theta_j}^2 \rho_{\ast}(r_h) c^3}{L_j(t-r_h/(\beta_j c))} \right]^{1/2}\right]^{-1},
\label{eq:betah}
\end{equation}
where $L_j(t-r_h/(\beta_j c))$ and $r_h$ refer to the jet luminosity
and the position of the jet head, respectively.  We use the formula
$L_j = \pi (r_h \theta_j)^2 \rho_j c^2 h_{j} \Gamma_{j}^2 \beta_j c$
in calculating Eq. \eqref{eq:betah}.  Accordingly, the position of the
jet head is calculated as $r_h(t) = \int_0^t \beta_h(t') c dt'$.

The jet head consists of shocked stellar matter and shocked jet
material. They are relativistically hot and expand sideways of the jet
forming a cocoon. We assume that almost all the jet energy goes
through the shocked region into the cocoon during the jet propagating
in the stellar envelope. The cocoon expands laterally by balancing its
pressure with the ram pressure of the stellar matter as
\begin{equation}
P_c = \rho_{\ast} c^2 h_{\ast}  \beta_c^2 + P_{\ast} \sim \rho_{\ast} c^2 \beta_c^2,
\label{eq:cocoonbalance}
\end{equation}
where the subscript c refers to the cocoon and $P_c\gg P_\ast$ is
assumed.  Since the cocoon consists of relativistically hot materials,
$P_c$ can be expressed as $P_c = E_c / (3V_c)$, using the cocoon
volume $V_c$ and the cocoon energy $E_c$. Now, we suppose the shape of
the cocoon as a cone, then $V_c(t) = \pi r_c^2(t) r_h(t)/3$.  In our
jet model, the cocoon energy can be expressed as $E_c(t) = \eta M_{\rm
  acc}(t) c^2$, where $M_{\rm acc}(t)$ is the mass accreted to the
black hole by the time $t$.  Substituting all these expressions into
Eq. \eqref{eq:cocoonbalance}, the cocoon expansion velocity $\beta_c$
can be calculated as
\begin{equation}
\beta_c(t) \sim \frac{r_h(t)}{r_c(t)} \sqrt{\frac{4 \eta M_{\rm acc}(t)}{3M(r_h)}},
\label{eq:betac}
\end{equation}
where $M(r_h) = (4 \pi r_h^3 /3) \rho_{\ast}$. Then the position of
the cocoon edge is given as $r_c(t) = \int_0^t \beta_c(t') c dt'$.

Now we discuss whether Pop III stars can raise GRBs by following the
time evolution of the positions of the jet head and the cocoon
edge. If the jet head reaches the stellar surface earlier than the
cocoon edge, we consider that the star can raise a GRB since we can
expect a successful jet breakout.  On the other hand, if the cocoon
edge reaches the stellar surface earlier, we can expect that the mass
accretion is suppressed and that the relativistic jet is stalled on
the way. This looks like a failed GRB.

Fig. \ref{fig:vel} shows the time evolution of the jet head velocity
$\beta_h$ (the red solid line) and the cocoon velocity $\beta_c$ (the
green dashed line).  In this figure, the time variability of
velocities comes from the discontinuity of the stellar density profile
and the mass accretion rate.  As pointed out in e.g.,
\cite{2001ApJ...556L..37M}, we can see that the jet head accelerates
drastically after entering the hydrogen envelope.  In addition, we
find that the jet head propagates faster than the cocoon edge all the
way through the stellar envelope except for the very early time. We
also find that the jet head breaks out of the stellar envelope $\sim
400$ sec after the central engine is activated.  Thus, we conclude
that a $40 M_\odot$ Pop III star has the possibility to raise a GRB.
 
Note here that $40M_{\odot}$ Pop III stars are thought to end their
lives as blue super giants (BSG), keeping large hydrogen envelopes
with radii $\sim 10^{12}$ cm. This is because the opacity is too low
to induce the mass loss from metal free stellar envelopes
\citep{2002RvMP...74.1015W}.  In general, the progenitor of a local
long GRB is not considered to be a super giant star with a hydrogen or
helium envelope but to be a Wolf-Rayet star with no hydrogen or helium
envelope and radius $\sim 10^{10}$ cm. The observational reason is
that every supernovae associating with long GRBs belongs to type
Ibc. Theoretically, it is considered that a super giant star has a too
largely extended envelope for the jet to break out successfully
\citep{2003MNRAS.345..575M} and it cannot raise a GRB.  From the
results here, however, we confirm that a BSG ($\sim 10^{12}$ cm) is
compact enough for a successful jet breakout.

We should note that \cite{2001ApJ...556L..37M} and \cite{2012ApJ...752...32W} 
have suggested the possibility of GRBs or gamma-ray transients from BSGs. However the
former treated a steady jet and did not quantitatively evaluate the
possibility, while the latter focused on the formation of the
accretion disk around the BH and did not discuss the jet break out. In
this section we quantitatively confirm the possibility of GRBs from
BSGs by consistently considering the stellar collapse, non-steady jet
injection, and the jet propagation.

\section{Observational Properties of Pop III GRBs}
\subsection{The prompt emission}
In this section, we consider observational characters of Pop III
GRBs. We assume that soon after the jet breakout, the jet emission can
be seen as a GRB and the burst lasts until the whole stellar envelope
accretes completely. We suppose that the efficiency for converting the
jet energy to the radiation energy is 10 \%. Accordingly, we can
calculate expected properties of the burst, such as the peak
luminosity ($L_{\rm p}$), duration ($T_{\rm 90}$) and the isotropic
energy ($E_{\gamma, {\rm iso}}$). Note here that we estimate the
$T_{\rm 90}$ as the period during which 90 \% of the burst's energy is
emitted.  Furthermore, we evaluate the time-integrated spectrum using
empirical laws for GRBs. There are some correlations which hold
between the time-integrated spectral peak energy in the observer frame
($E_p^{\rm obs}$) and the peak luminosity ($L_{\rm p}$) or the
isotropic energy ($E_{\gamma,{\rm iso}}$). One is the $E_p-L_p$
correlation \citep{2004ApJ...609..935Y} and the other is the
$E_p-E_{\gamma,{\rm iso}}$ correlation \citep{2002AA...390...81A}. The
functional forms of these two correlations are represented as

\begin{equation}
\frac{L_{p}}{10^{52}\ {\rm erg} \ {\rm sec}^{-1}} \sim 2 \times 10^{-5}\left[\frac{E_p^{\rm obs} (1+z)}{1\ {\rm keV}} \right]^{2.0},
\label{eq:Yonetoku}
\end{equation}

\begin{equation}
\left[\frac{E_p^{\rm obs} (1+z)}{1\ {\rm keV}} \right] \sim 80 \left[\frac{E_{\rm \gamma,iso}}{10^{52}\ {\rm erg}} \right]^{0.57},
\label{eq:Amati}
\end{equation}
respectively.

Table \ref{tab:popIIIGRB} shows the results of our model for a
$40M_{\odot}$ Pop III star. As can be seen from Table
\ref{tab:popIIIGRB}, Pop III GRBs radiate as much energy ($E_{\gamma,
  \rm iso} \sim 10^{54}$ erg) as the most energetic local long GRBs
do, while $L_p \sim 5 \times 10^{50} \ {\rm erg} \ {\rm sec}^{-1}$ is
smaller by a factor of $\sim 10$. Moreover, the duration ($\sim
10^{5}$ sec) of Pop III GRBs is much longer than that of local long
GRBs.\footnote{In \cite{2011ApJ...726..107S}, they evaluated the
  duration of the Pop III GRB from a $915M_{\odot}$ star as $\sim
  1,500$ sec in the GRB frame, but this is a wrong value. We find that
  the correct value is $\sim 15,000$ sec in the GRB frame, which is
  similar to the one obtained here ($\sim 6000$ sec in the GRB
  frame). This is because the larger mass of the $915M_{\odot}$ Pop
  III star compensates with its larger radius ($\sim 10^{13}$ cm), as
  we can see from the expression of the free-fall time.}  All these
differences come from the fact that although the progenitor of a local
long GRB has no hydrogen envelope and is more compact, a Pop III
progenitor has a large hydrogen envelope. Since a Pop III star
experiences no mass loss and keeps a more massive hydrogen envelope,
the energy supply to the central engine can last longer time.  This
enables the central engine to be kept active for much longer
time. This also enables the burst to have a much longer duration and
to have the vast isotropic energy. On the other hand, it takes longer
time for the jet to break out the larger stellar envelope and the jet
energy after the breakout is more damped. This causes the Pop III GRBs
to have lower luminosities.

We evaluate the observed peak energy for either the case that the
$E_p-L_p$ correlation holds or that the $E_p-E_{\gamma, \rm iso}$
correlation does. For the case of $E_p-L_p$ correlation, $E_p^{\rm
  obs} \sim 5$ keV is in the X-ray region, whereas for the
$E_p-E_{\gamma, \rm iso}$ correlation, the peak energy in the GRB
frame is larger than that of a local long GRB, because of the larger
$E_{\gamma, \rm iso}$ value. However, the cosmological redshift effect
reduces the peak down to $E_p^{\rm obs} \sim 120$ keV, which is
slightly softer than that of a local long GRB.

\subsection{The detectability of Pop III GRBs}
In this subsection, we discuss the detectability of Pop III GRBs. In
\cite{2011ApJ...726..107S}, they found that Pop III GRBs are too dim
to trigger {\it Swift} Burst Alert Telescope (BAT).  We obtain the
similar conclusion even we employ the different progenitor from them.
Therefore, we here discuss whether Pop III GRBs trigger the future
satellite missions such as {\it Lobster} \citep{2012IAUS..285...41G}
and {\it EXIST} \footnote{http://exist.gsfc.nasa.gov/} in
detail. While {\it Lobster} will have energy window range of $0.3-5$
keV, {\it EXIST} will have that of $5-600$ keV.
  
An event is regarded to be detected if the number of the signal
photons within the detector energy range $[E_{\rm min}, E_{\rm max}]$
satisfies the following relation,
\begin{equation}
\frac{\int_{t_0}^{t_0+\Delta t} N_{\rm sig}({t}_{\rm obs}') d{t}_{\rm obs}' A }{\left(\int_{t_0}^{t_0+\Delta t} N_{\rm bg} d{t}_{\rm obs}' A \right)^{1/2}} \gtrsim ({\rm S/N})_{\rm min}.
\label{eq:SN}
\end{equation}
Here, $t_0$ is the time when an event comes in the detector's field of
view and the detector starts to observe the event, $\Delta t$ is the
exposure time for the event, and $t_{\rm obs}$ is the time from the
beginning of the burst in the observer frame. $N_{\rm sig}, A$ and
$N_{\rm bg}$ refer to the signal photon number flux, the area of the
detector and the number flux of background photons,
respectively. $({\rm S/N})_{\rm min}$ is the critical signal to noise
ratio needed for detection.  Assuming that the background photon flux
is constant and using the signal photon energy flux within the
detector energy window range $[E_{\rm min}, E_{\rm max}]$,
Eq. \eqref{eq:SN} can be written as
\begin{equation}
\Bar{f}_{\rm sig}(t_0, \Delta t) \gtrsim f_{\rm sen}(\Delta t), 
\label{eq:crit1}
\end{equation}
where 
\begin{equation}
\Bar{f}_{\rm sig}(t_0, \Delta t) \equiv \frac{\int_{t_0}^{t_0+\Delta t} f_{\rm sig}({t}_{\rm obs}') d{t}_{\rm obs}' }{\Delta t}
\label{eq:aveflux}
\end{equation}
is the energy flux averaged over the exposure time and
\begin{equation}
f_{\rm sen}(\Delta t) \equiv \frac{\int_{t_0}^{t_0+\Delta t} f_{\rm sig} d{t}_{\rm obs}' }{\int_{t_0}^{t_0+\Delta t} N_{\rm sig} d{t}_{\rm obs}' } ({\rm S/N})_{\rm min} A^{-1/2} N_{\rm bg}^{1/2} {\Delta t}^{-1/2}
\label{eq:sense}
\end{equation}
is the energy flux sensitivity within $[E_{\rm min}, E_{\rm max}]$.
In our model, $f_{\rm sig}({t}_{\rm obs})$ can be calculated from
\begin{equation}
f_{\rm sig}({t}_{\rm obs}) = \frac{L_{\gamma, \rm iso}({t}_{\rm obs})}{4 \pi d_L^2} \frac{\int_{E_{\rm min}}^{E_{\rm max}} E N(E) dE}{\int_{0}^{\infty} E N(E) dE}\ {\rm erg}\ {\rm cm}^{-2}\ {\rm sec}^{-1}.
\label{eq:flux}
\end{equation}
In Eq. \eqref{eq:flux}, $L_{\gamma, \rm iso}({t}_{\rm obs})$ is the
isotropic equivalent luminosity of the burst at ${t}_{\rm obs}$. $d_L$
is the luminosity distance calculated with cosmological parameters
$(\Omega_m, \Omega_{\Lambda}) = (0.28, 0.72)$ and the Hubble parameter
$H_0 = 70\ {\rm km}\ {\rm sec}^{-1}\ {\rm Mpc}^{-1}$. $N(E)$ is the
Band spectrum \citep{1993ApJ...413..281B} with the typical parameter
values, $\alpha = -1$ and $\beta = -2.3$.  We discuss the
detectability of Pop III GRBs for either the case that the $E_p-L_p$
correlation holds or that the $E_p-E_{\gamma, \rm iso}$ correlation
does.

First, we consider the case of the $E_p-L_p$ correlation. Since
$E_p^{\rm obs} \sim 5$ keV, in this case {\it Lobster} is more
appropriate for detecting Pop III GRBs. Recently,
\cite{2010A&A...511A..43G} studied the time dependent spectral
characteristics of several individual bright GRBs. They found that the
isotropic equivalent luminosity $L_{\gamma, \rm iso}({t}_{\rm obs})$
correlates with the time resolved spectrum peak energy $E_p({t}_{\rm
  obs})$ for each GRB and that the functional form of the correlation
is very similar to the time integrated $E_p-L_p$ correlation
(Eq. \eqref{eq:Yonetoku}).  Note that they calculated the time
resolved spectrum by integrating the signal flux within 1 sec time bin
around each time. Accordingly, if we assume the validity of the
time-resolved $E_p({t}_{\rm obs})-L_{\gamma, \rm iso}({t}_{\rm obs})$
correlation, which is obtained by replacing $L_p$ and $E_p^{\rm obs}$
in Eq. \eqref{eq:Yonetoku} by $L_{\gamma, \rm iso}({t}_{\rm obs})$ and
$E_p^{\rm obs}({t}_{\rm obs})$, we can discuss the detectability using
the condition in Eq. \eqref{eq:crit1}.

The {\it Lobster} sensitivity for a soft source (a power-law photon
index of $-2$) is estimated to be $1.3 \times 10^{-11}$ erg\ ${\rm
  cm}^{-2}$\ ${\rm sec}^{-1}$ ($0.3-5$ keV, $5 \sigma$) at one calendar
day (an effective exposure time of $\sim 2500$ sec; see
\citealp{2012IAUS..285...41G}).  On the other hand, the sensitivity
for a proposed exposure time per pointing in a realistic operation
($\sim 450$ sec) is calculated to be $3.1 \times 10^{-11}$ erg\ ${\rm
  cm}^{-2}$\ ${\rm sec}^{-1}$ ($0.3-5$ keV, $5 \sigma$).  The assumed
spectral parameter in this estimation is reasonable for a GRB with
$E_p^{\rm obs} \sim$ a few keV.  We discuss the detectability of a Pop
III GRB by {\it Lobster} using the sensitivity in $\Delta t \sim 450$
sec as a realistic case and in $\Delta t \sim 2500$ sec as an
optimistic case.

\begin{figure}[!htb]
\includegraphics[angle=-90,scale=1.0]{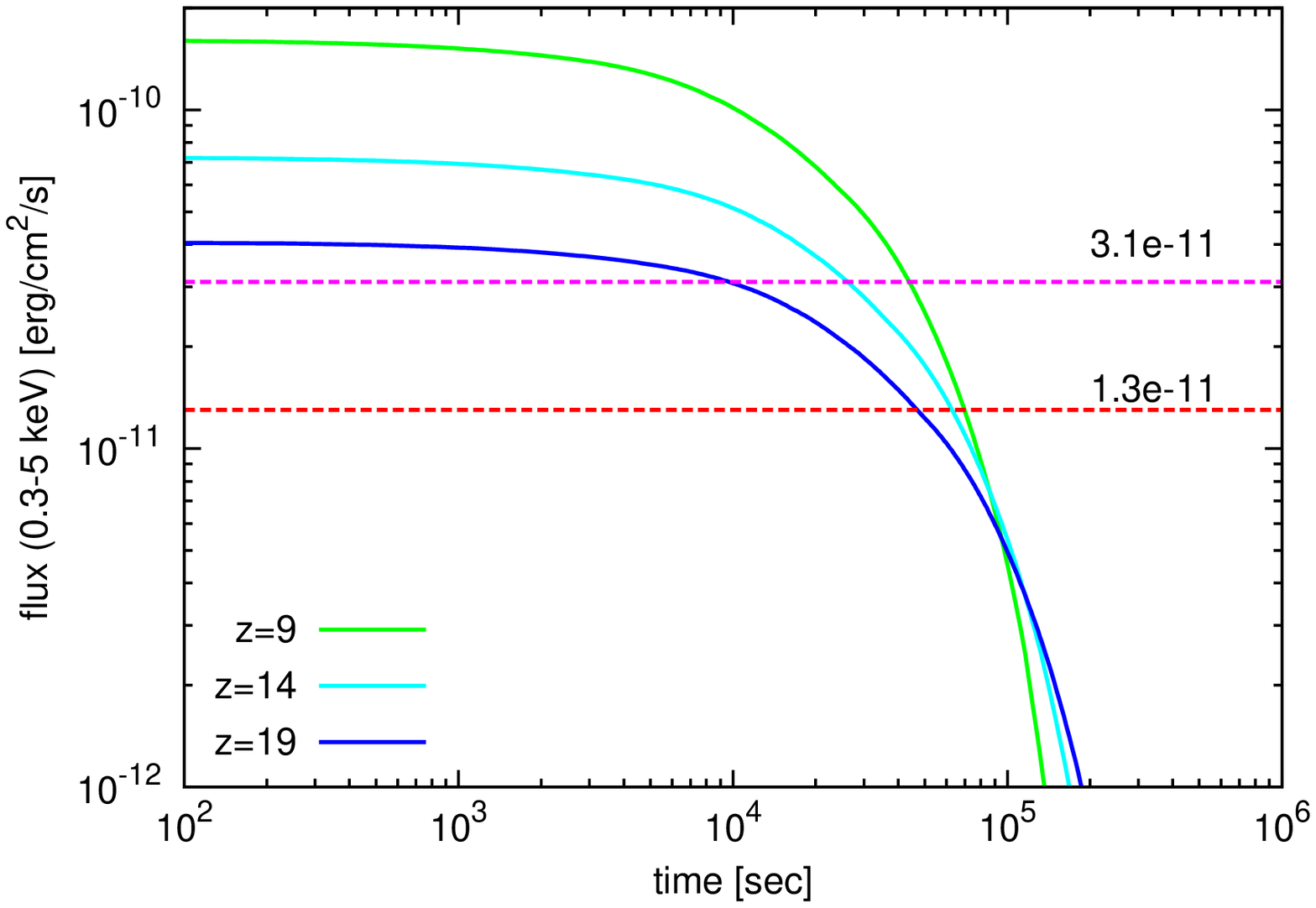}
\caption{The comparison of the energy flux $f_{\rm sig}(t_0)$ for a
  Pop III GRB with the detection thresholds of {\it Lobster}, $f_{\rm
    sen}(\Delta t)$ (0.3-5 keV, $5 \sigma$). The abscissa is the time
  from the beginning of a GRB, i.e., from the jet break out, in the
  observer frame. $t_0$ is the time when the event comes into the {\it
    Lobster} field of view and {\it Lobster} starts to observe the
  event. $\Delta t$ is the proposed exposure time of {\it Lobster}.
  The green, sky-blue and the blue solid lines correspond to ${f}_{\rm
    sig}$ (0.3-5 keV) for the GRB from a $40M_{\odot}$ Pop III star at
  $z=9, 14$ and $19$, respectively. The red and the magenta dashed
  lines represent $f_{\rm sen}(\Delta t)$ (0.3-5 keV, $5 \sigma$) of
  {\it Lobster} \citep{2012IAUS..285...41G}, corresponding to the
  optimistic case of $\Delta t \sim 2500$ sec and the realistic case
  of $\Delta t \sim 450$ sec, respectively. If $f_{\rm sig}(t_0)
  \gtrsim f_{\rm sen}(\Delta t)$ holds, we regard that {\it Lobster}
  observes the Pop III GRB from $t_0$ to $t_0 + \Delta t$.}
\label{fig:LC}
\end{figure}

In Fig. \ref{fig:LC}, we compare the energy flux of GRBs from $40
M_{\odot}$ Pop III stars with the detection thresholds of {\it
  Lobster}. The abscissa is the time from the beginning of a GRB,
i.e. from the jet break out, in the observer frame. The green,
sky-blue and the blue solid lines represent $f_{\rm sig}({t}_{\rm
  obs})$ of Pop III GRBs at $z = 9, 14$ and $19$, respectively,
calculated from Eq. \eqref{eq:flux}. The red and magenta dashed lines
correspond to $f_{\rm sen}$ of {\it Lobster} in a realistic case
(magenta) and an optimistic case (red).  From Fig. \ref{fig:LC}, we
can see that $f_{\rm sig}({t}_{\rm obs})$ does not change
significantly over $\Delta t \sim 450$ sec or $\sim 2500$ sec around
each time, so we can approximate $f_{\rm sig} \sim$ const. over the
considered exposure times. Then, Eq. \eqref{eq:crit1} can be rewritten
as
\begin{equation}
{f}_{\rm sig}(t_0) \gtrsim f_{\rm sen}(\Delta t). 
\label{eq:criLob}
\end{equation}
Eq. \eqref{eq:criLob} indicates that if ${f}_{\rm sig}(t_0)$ when an
event comes into the {\it Lobster} field of view is larger than
$f_{\rm sen}(\Delta t)$ for given $\Delta t$, we can observe the event
from $t_0$ to $t_0 + \Delta t$.  From Fig. \ref{fig:LC}, we find that
Pop III GRBs at $z =9, 14$ and even at $z =19$ have the possibility to
trigger {\it Lobster}. {\it Lobster} will detect a Pop III GRB as a
long duration X-ray flash with nearly constant luminosity.

Subsequently, we consider the case of the $E_p-E_{\gamma, \rm iso}$
correlation.  Since $E_p^{\rm obs} \sim 120$ keV, in this case {\it
  EXIST} is the better instrument for detection.  Note that
the $E_p-E_{\gamma, \rm iso}$ correlation is the correlation between
the total radiated energy and the time-integrated spectrum, we can
regard $E_p^{\rm obs}$ as the observed peak energy time-averaged
within each burst. So, we evaluate $f_{\rm sig}$ ($5-600$ keV)
assuming that the spectrum is the Band type with $E_p^{\rm obs} \sim
120$ keV, $\alpha=-1$ and $\beta=-2.3$ and that the spectrum does not
change with time.  The sensitivity of {\it EXIST} for a proposed
exposure time in the longest time-scale at the on-board process
($\Delta t \sim 512$ sec) is calculated to be $f_{\rm sen} \sim 2.4
\times 10^{-10}\ {\rm erg}\ {\rm cm}^{-2}\ {\rm sec}^{-1}$ ($5-600$
keV, $5 \sigma$) \citep{2009SPIE.7435E...6H}.

\begin{figure}[!htb]
\includegraphics[angle=-90,scale=1.0]{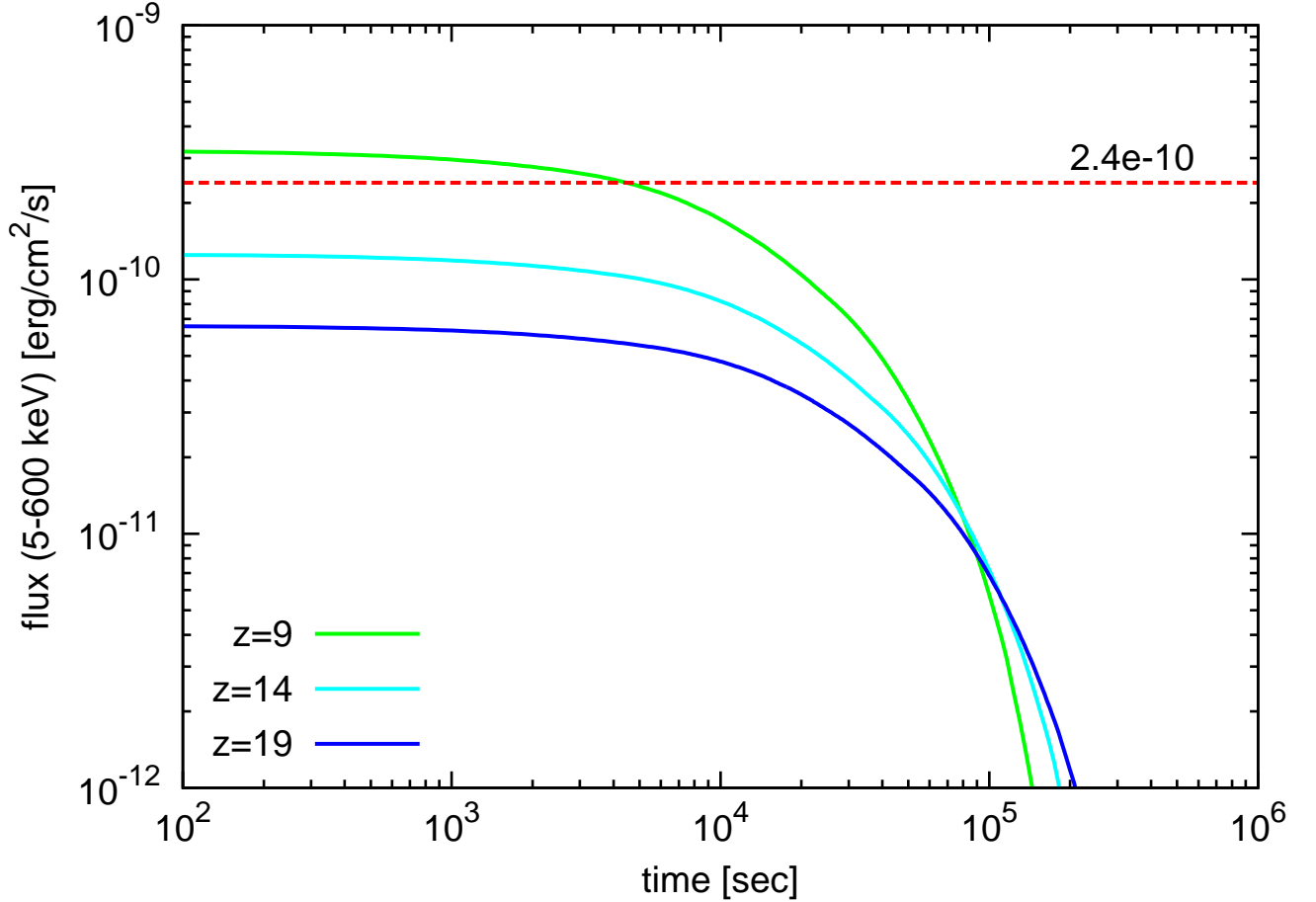}
\caption{Same as Fig. \ref{fig:LC} but for {\it EXIST} case. {\it
    EXIST} will have the limited energy range of 5-600 keV.  The red
  dashed line represents the {\it EXIST} sensitivity $f_{\rm sen} \sim
  2.4 \times 10^{-10}\ {\rm erg}\ {\rm cm}^{-2}\ {\rm sec}^{-1}$
  (5-600 keV, $5 \sigma$) in the longest exposure time-scale at the
  on-board process ($\Delta t \sim 512$ sec)
  \citep{2009SPIE.7435E...6H}.}
\label{Fig:EXIST}
\end{figure}

We show the results for $40 M_{\odot}$ Pop III stars in
Fig. \ref{Fig:EXIST}. Here again, the abscissa is the time from the
beginning of a burst, i.e. from the jet break out, in the observer
frame. The green, sky-blue and the blue solid lines represent $f_{\rm
  sig}({t}_{\rm obs})$ of Pop III GRBs at $z = 9, 14$ and $19$,
respectively, calculated from Eq. \eqref{eq:flux}. The red dashed line
represents the {\it EXIST} sensitivity described above. In this case
also, $f_{\rm sig}({t}_{\rm obs})$ is approximately constant over
$\Delta t \sim 512$ sec, so Eq. \eqref{eq:crit1} can be rewritten in
the form Eq. \eqref{eq:criLob}. From Fig. \ref{Fig:EXIST}, we can see
that although Pop III GRBs at $z = 14$ and $19$ do not trigger {\it
  EXIST}, Pop III GRBs at $z=9$ have the possibility to trigger {\it
  EXIST}. {\it EXIST} will detect such a Pop III GRB as a long
duration X-ray rich GRB with nearly constant luminosity.

\subsection{Other progenitor models}
\cite{2008ApJ...681..771M} and \cite{2011Sci...334.1250H} studied the
mass of a Pop III star at its birth by calculating the evolution of a
primordial protostar in analytical or numerical way. Because the
initial angular momentum of a primordial gas cloud is considered to be
large enough, in the star formation phase, a protostar and a
circumstellar accretion disk system is formed and the protostar gains
mass by the accretion of the surrounding gas through the disk. They
found that the UV radiation from the protostar eventually stops the
mass accretion and the growth of the star by evaporating the
surrounding gas and eventually the disk. They also found that the
final mass of a Pop III star depends on the degree of the angular
momentum transport within the accretion disk and on the magnitude of
the initial angular momentum of a gas cloud. In
\cite{2011Sci...334.1250H}, the degree of the angular momentum
transport is characterized by the $\alpha_0$-parameter of the disk,
where the larger $\alpha_0$ value means the larger mass accretion. In
fig. S1 of \cite{2011Sci...334.1250H}, a Pop III star finally gains
$\sim 50M_{\odot}$ with $\alpha_0 = 1.0$, $\sim 40M_{\odot}$ with
$\alpha_0 = 0.6$ (fiducial case), and $\sim 35M_{\odot}$ with
$\alpha_0 = 0.3$, for the fiducial magnitude of the initial angular
momentum. Furthermore, the mass of a Pop III star depends on the
initial angular momentum of the star-forming gas cloud and it gets
$\sim 85M_{\odot}$ when the initial angular momentum is reduced by 30
\% of the fiducial one with $\alpha_0 = 0.6$ (fiducial case).

\begin{figure}[!htb]
\includegraphics[angle=-90,scale=1.0]{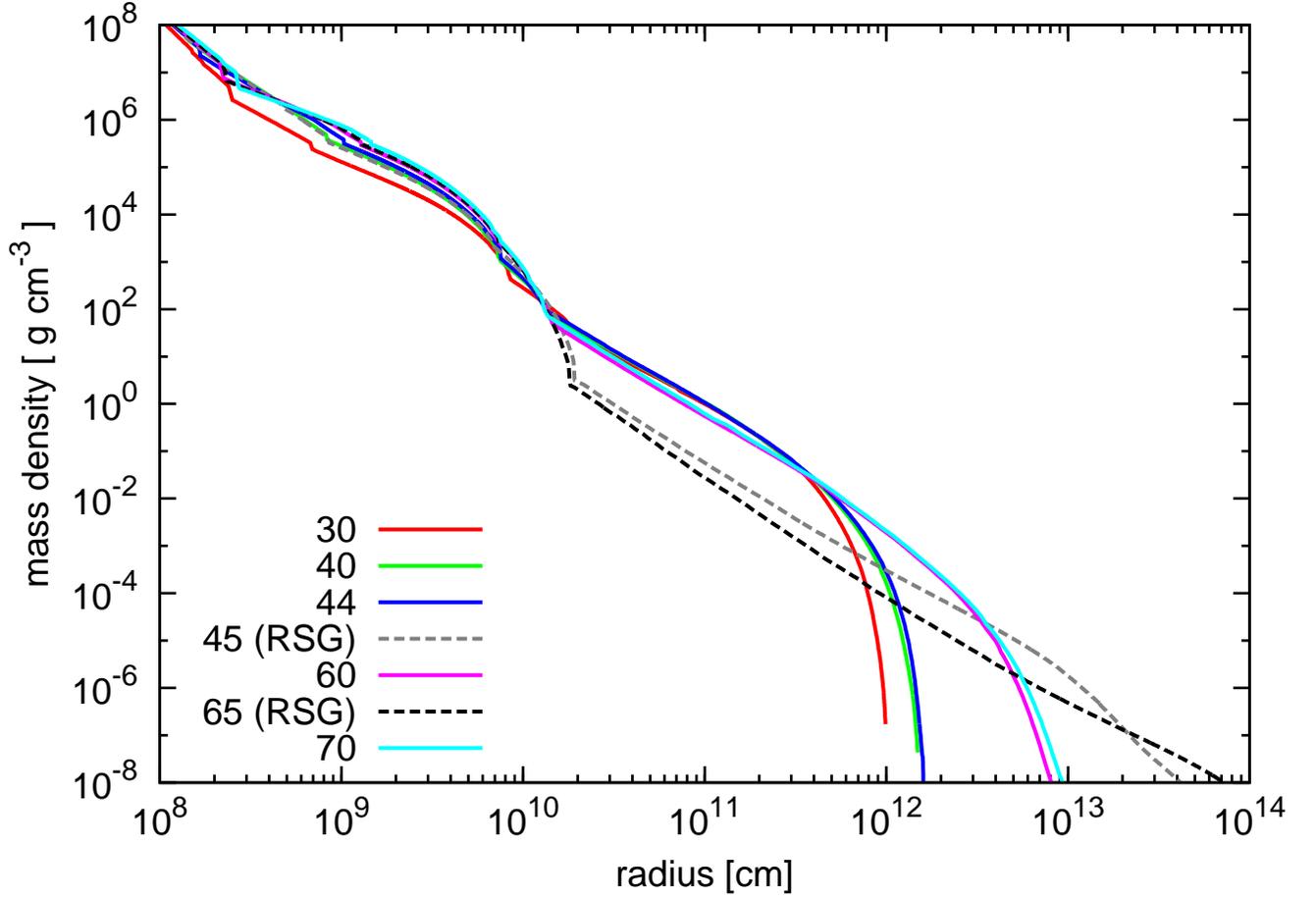}
\caption{The density profiles for some of the Pop III stellar models
  \citep{2002RvMP...74.1015W, 2010ApJ...724..341H}. Those who end
  their life as BSGs correspond to $30 M_{\odot}$ (red), $40
  M_{\odot}$ (green), $44 M_{\odot}$ (blue), $60 M_{\odot}$ (magenta)
  and $70 M_{\odot}$ (sky-blue), respectively. They have radii
  $10^{12-13}$ cm. Those who end their life as RSGs correspond to $45
  M_{\odot}$ (grey) and $65 M_{\odot}$ (black), respectively. They
  have radii $\sim 10^{14}$ cm. We can see that RSGs have much more
  extended and lower density envelopes than BSGs have.}
\label{fig:prof}
\end{figure}

Accordingly, in this subsection, we investigate whether $30 - 90
M_{\odot}$ Pop III stars can be the progenitors of GRBs. The stellar
models of $30 - 40 M_{\odot}$ are given by \cite{2002RvMP...74.1015W}
and those of $41 - 90M_{\odot}$ are from
\cite{2010ApJ...724..341H}. In Fig. \ref{fig:prof}, we show the
density profiles of selected models. \cite{2002RvMP...74.1015W} showed
that all the $30 - 40M_{\odot}$ Pop III stars end their lives as blue
super giants (BSG). According to \cite{2010ApJ...724..341H}, although
$41 - 44, 60$ and $70M_{\odot}$ Pop III stars end as BSGs, $45, 50,
55, 65, 75, 80, 85$ and $90M_{\odot}$ Pop III stars end as red super
giants (RSG). As shown in \cite{2010ApJ...724..341H}, the RSG branch
in the higher mass stars appears due to the primary nitrogen
production in the hydrogen burning shell.

For these stellar models, we investigate whether these stars can raise
GRBs by considering the jet propagation in the stellar envelope with
the entirely similar manner as in \S 2. Note here that although for a
Pop III star with $M \lesssim 40M_\odot$ the fall back effect should
be taken into consideration in the formation of a black hole remnant
(see e.g., \citealp{2001ApJ...550..410M,2008MNRAS.388.1729K}), we
neglect this effect in this section and discuss it in \S4.  For BSGs
($30 - 44, 60$ and $70M_{\odot}$), the jet head propagates faster than
the cocoon edge almost all the way through the stellar envelope like
 in Fig. \ref{fig:vel}. On the other hand, for RSGs, we find that
the jet head reaches the surface as early as or even later than the
cocoon edge does. Therefore, we conclude that although Pop III BSGs ($30 -
44, 60, 70M_{\odot}$) have the possibility to raise GRBs, Pop III RSGs
do not. From the above discussions, we can say that although Pop III
BSGs ($\sim 10^{12}$ cm) are compact enough for successful jet
breakouts, Pop III RSGs ($\sim 10^{14}$ cm) have too largely extended
low density envelopes for jets to break out successfully
(Fig. \ref{fig:prof}).

\begin{figure}[!htb]
\includegraphics[angle=-90,scale=1.0]{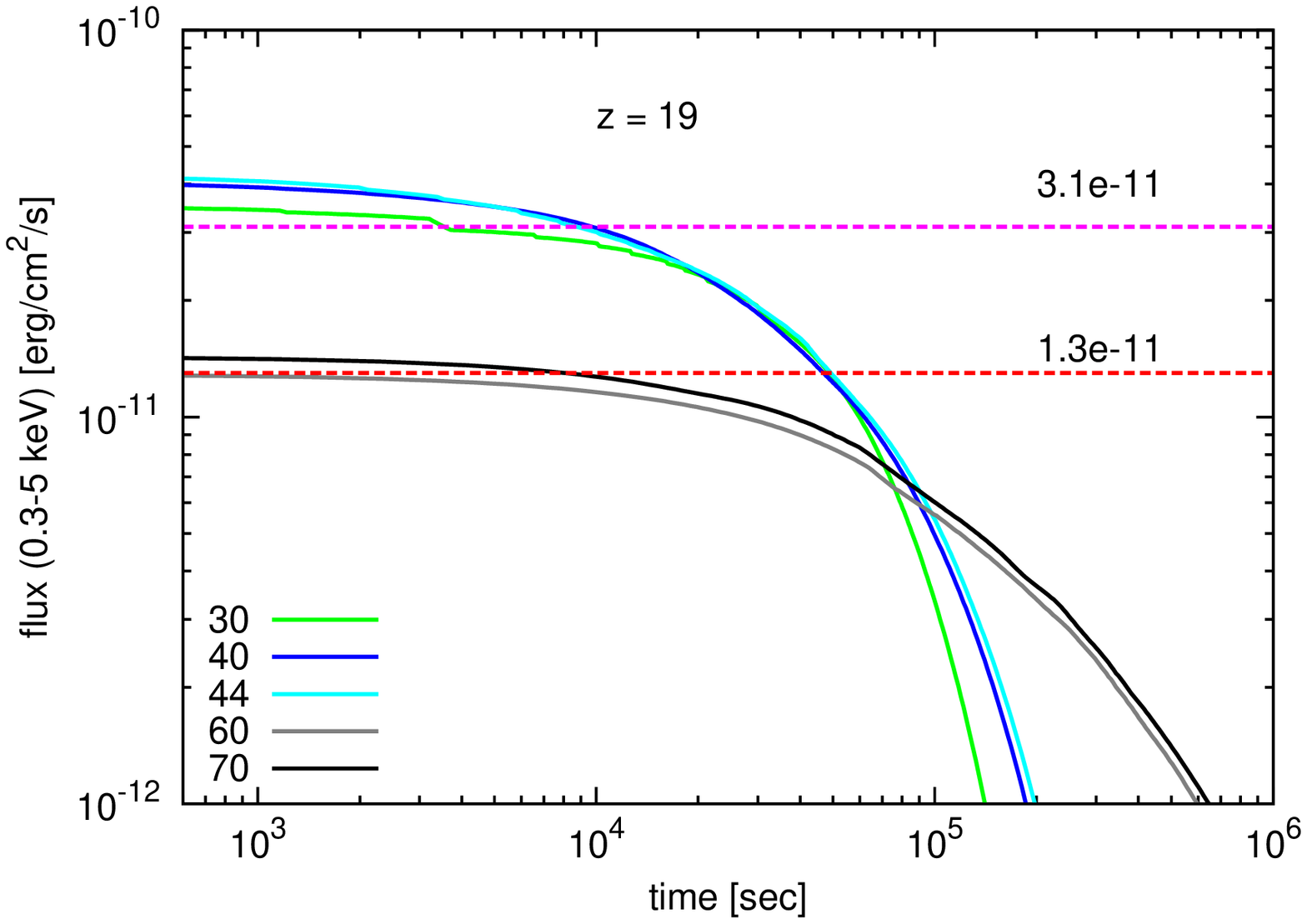}
\caption{The comparison of the energy flux $f_{\rm sig}(t_0)$ for Pop
  III GRBs with the detection thresholds of {\it Lobster}, $f_{\rm
    sen}(\Delta t)$ (0.3-5 keV, $5 \sigma$). The abscissa is the time
  from the beginning of a GRB, i.e., from the jet break out, in the
  observer frame. $t_0$ is the time when the event comes into the {\it
    Lobster} field of view and {\it Lobster} starts to observe the
  event. $\Delta t$ is the proposed exposure time.  The green, blue,
  sky-blue, grey and black solid lines correspond to ${f}_{\rm sig}$
  (0.3-5 keV) of Pop III GRBs at $z=19$ from $30, 40, 44, 60$ and
  $70M_{\odot}$ progenitors, respectively. The red and the magenta
  dashed lines represent $f_{\rm sen}(\Delta t)$ (0.3-5 keV, $5
  \sigma$) of {\it Lobster} \citep{2012IAUS..285...41G}, corresponding
  to the optimistic case of $\Delta t \sim 2500$ sec and the realistic
  case of $\Delta t \sim 450$ sec, respectively. If $f_{\rm sig}(t_0)
  \gtrsim f_{\rm sen}(\Delta t)$ holds, we regard that {\it Lobster}
  observes the Pop III GRB from $t_0$ to $t_0 + \Delta t$.}
\label{fig:Lobmass}
\end{figure}

\begin{figure}[!htb]
\includegraphics[angle=-90,scale=1.0]{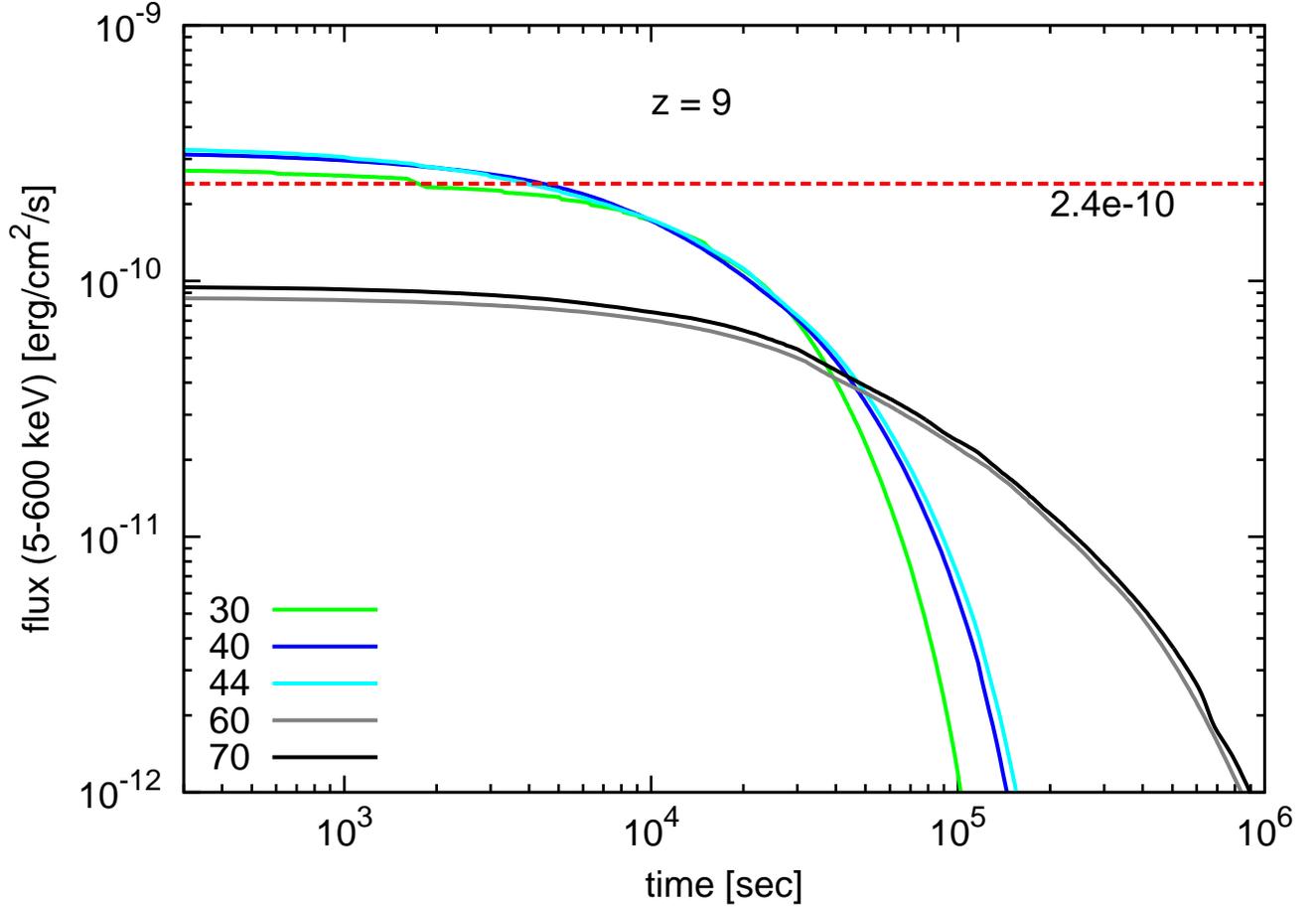}
\caption{Same as Fig. \ref{fig:Lobmass}, but for {\it EXIST} (5-600
  keV) case. The red dashed line represents the {\it EXIST}
  sensitivity $f_{\rm sen} \sim 2.4 \times 10^{-10}\ {\rm erg}\ {\rm
    cm}^{-2}\ {\rm sec}^{-1}$ (5-600 keV, $5 \sigma$) in the longest
  exposure time-scale at the on-board process ($\Delta t \sim 512$
  sec) \citep{2009SPIE.7435E...6H}. Note that we focus on Pop III GRBs
  at $z = 9$ in this figure.}
\label{fig:EXISTmass}
\end{figure}

Subsequently, we evaluate the observational characters and the
detectability of these Pop III GRBs at $z=19$. The results for some
models are shown in Table \ref{tab:obsmass}. We can see the same
characteristics for Pop III GRBs as described in \S 3.1.
Here again, we consider either the case that the $E_p-L_p$ correlation
holds or that the $E_p-E_{\gamma, \rm iso}$ correlation holds. First,
we consider the case of the $E_p-L_p$ correlation. In
Fig. \ref{fig:Lobmass}, the red and the magenta dashed lines represent
$f_{\rm sen}(\Delta t)$ (0.3-5 keV, $5 \sigma$) of {\it Lobster} in
the optimistic and realistic case, respectively
\citep{2012IAUS..285...41G}. The green, blue, sky-blue, grey and black
solid lines correspond to the energy flux $f_{\rm sig}({t}_0)$ of Pop
III GRBs at $z =19$ for $30, 40, 44, 60$ and $70M_{\odot}$
progenitors, respectively.  The abscissa is the time from the
beginning of each burst, i.e. from the jet break out. We find that
while it is difficult for Pop III GRBs from 60 and 70 $M_{\odot}$
progenitors to trigger {\it Lobster}, Pop III GRBs from $\lesssim 44
M_{\odot}$ stars have the possibility to trigger {\it Lobster}.
Second, we consider the case of the $E_p-E_{\gamma, \rm iso}$
correlation and Fig. \ref{fig:EXISTmass} shows the results. The red
dashed line represents the {\it EXIST} sensitivity.
Here also, the green, blue, sky-blue, grey and black solid lines
correspond to the energy flux $f_{\rm sig}({t}_0)$ of Pop III GRBs for
$30, 40, 44, 60$ and $70M_{\odot}$ progenitors, respectively, but at
$z =9$.
We find that 
only Pop III GRBs from $\lesssim 44 M_{\odot}$ stars have the
possibility to trigger {\it EXIST}.

From Fig. \ref{fig:Lobmass} and Fig. \ref{fig:EXISTmass}, we can see
that the energy fluxes of Pop III GRBs from $\lesssim 44 M_{\odot}$
progenitors are larger by a factor of 3 to 4 than those from $60, 70
M_{\odot}$ stars. This is because more massive progenitors exhibit
larger radii (see Fig. \ref{fig:prof}). The larger radius a progenitor
has, the longer time it takes for the jet to reach the stellar surface
and the more damped the jet luminosity after the breakout is.
   
\subsection{The afterglow}

\begin{figure}[!htb]
\includegraphics[angle=-90,scale=1.0]{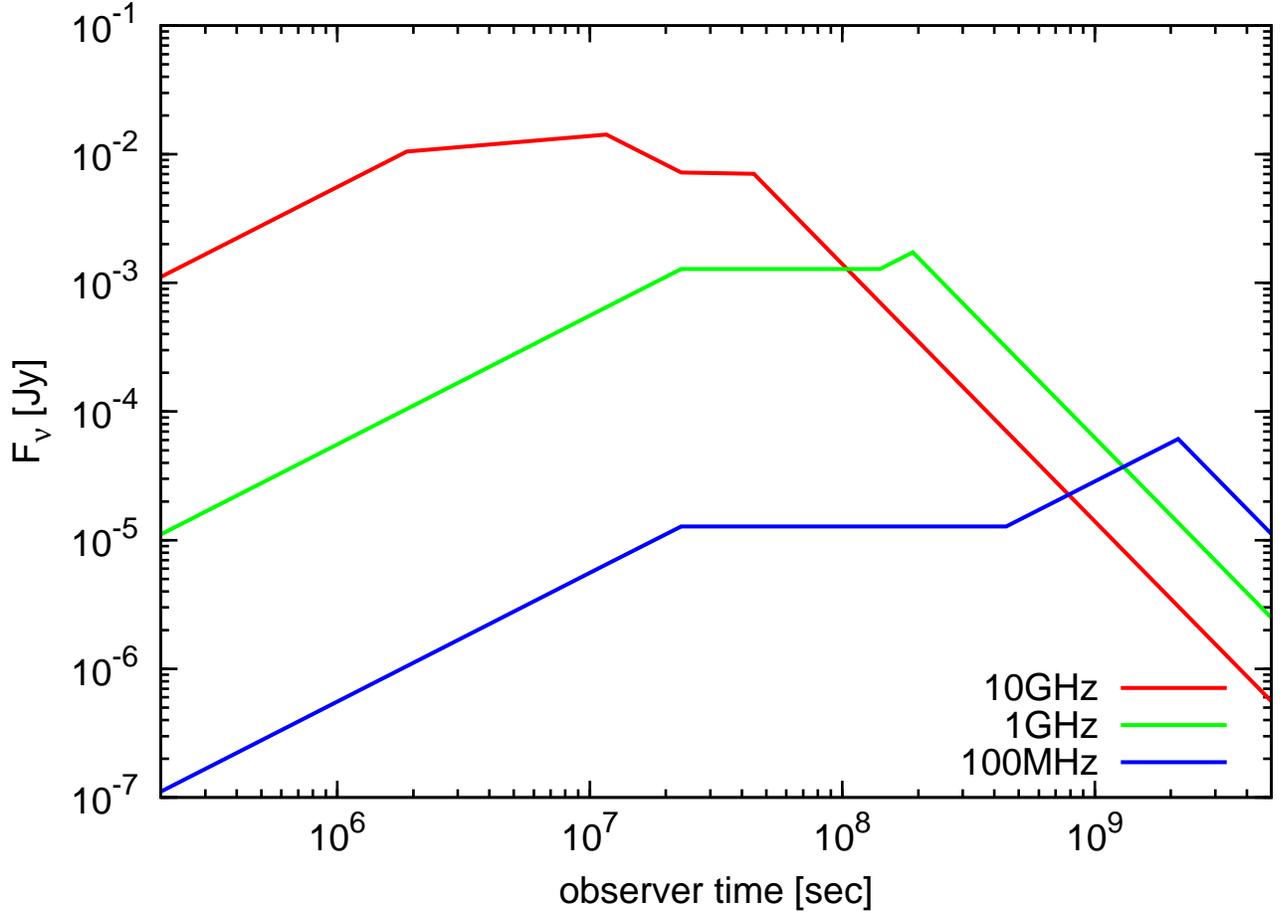}
\caption{The afterglow light-curves of a Pop III GRB at 100 MHz, 1 GHz
  and 10 GHz in the case that the GRB happens at $z=19$. The red,
  green and the blue solid lines correspond to the light-curves at 10
  GHz, 1 GHz and 100 MHz, respectively. The abscissa is the time from
  the beginning of the prompt emission, i.e. from the jet break out
  time, in the observer frame. Here, we adopt the parameter values of
  $n=E_{55}=\epsilon_{e, -1} = \epsilon_{B, -2} =f(p)=\theta_{j,
    -1}=1$ and $p=2.3$. $Q_{x}$ denotes $Q/10^{x}$.}
\label{fig:afg}
\end{figure}

In this subsection, let us discuss the afterglow of a Pop III GRB
following \cite{2011ApJ...731..127T}. In the external shock model of
an afterglow, we consider that a relativistic ejecta with isotropic
equivalent kinetic energy $E_{\rm iso}$ and a half opening angle
$\theta_j$ moves through the interstellar medium with density $n$
making a shocked region at the head of it. In the shocked region, some
fractions, $\epsilon_B$ and $\epsilon_e$, of the internal energy
are provided to the magnetic field energy and the energy for
the electron acceleration, respectively. The accelerated electrons are assumed to
have a number distribution of the form $N(\gamma_e) \propto
\gamma_e^{-p}$ and to raise afterglow emissions through the
synchrotron radiation and the inverse Compton emission. As fiducial
parameter values of the external shock model, we adopt $E_{\rm iso}
\sim 10^{55}$ erg, $n = 1$ ${\rm cm}^{-3}$, $\epsilon_e = 0.1,
\epsilon_B = 0.01, \theta_j=0.1$ and $ p=2.3$. In \S 3, we assumed
that 10 \% of the jet energy was converted into the energy for the
prompt photon emission. Accordingly, we consider that the remaining 90
\% of the jet energy is used for afterglow emissions. As we saw in
\S3, $E_{\gamma, \rm iso} \sim 10^{54}$ erg, so here we adopt $E_{\rm
  iso} \sim 10^{55}$ erg as a fiducial value. We refer to
\cite{2011ApJ...731..127T} for other parameter values.  Under this
model, we calculate the afterglow light-curves at 10 GHz, 1 GHz and
100 MHz. We show the results in Fig. \ref{fig:afg} in the case of a
Pop III GRB at $z=19$. The red, green and the blue solid lines refer
to the light curves at 10 GHz, 1 GHz and 100 MHz, respectively.

Let us discuss the detectability of such Pop III GRB radio afterglow
emissions. The Low Frequency Array
(LOFAR)\footnote{http://www.astron.nl/} has frequency coverage from 10
to 250 MHz and the detection threshold of 0.2 mJy ($1 \sigma$ level)
at 100 MHz for 1 hr integration time. From Fig. \ref{fig:afg}, we find
that the 100 MHz radio afterglow emission is not detectable by
LOFAR. On the other hand, the Expanded Very Large Array (EVLA) has
frequency coverage from 1 to 50 GHz and the detection thresholds of
5.5 $\mu$Jy and 1.8 $\mu$Jy ($1 \sigma$ level) at 1 GHz and 10 GHz,
respectively for 1 hr integration time \citep{2011ApJ...739L...1P}.
From Fig. \ref{fig:afg}, we can see that the energy fluxes at 1 GHz
and 10 GHz are much larger than the detection thresholds of EVLA. Once
a Pop III afterglow emerges, we have the possibility to detect it at
any time by EVLA, even if it occurs at such a high redshift universe
($z=19$).

\section{Summary and discussion}
GRBs are the brightest phenomena in the universe. Their detections at
high $z$ universe ($z \sim 9$) motivate us to expect GRBs to be one of
the powerful tools to probe the early universe. Focusing on the high
$z$ universe, we should consider the association of GRBs with Pop III
stars. Recent numerical simulations \citep{2011Sci...334.1250H}
suggest that Pop III stars obtain mass typically $\sim 40M_{\odot}$ at
their birth. Zero metallicity stars are considered not to lose mass
during entire life because of the low opacity envelopes
\citep{2002RvMP...74.1015W}. Therefore, they enter into the
pre-supernova stage keeping large hydrogen envelopes. According to
\cite{2002RvMP...74.1015W} and \cite{2010ApJ...724..341H}, Pop III
stars end their lives as BSGs or RSGs depending on the amount of
primary nitrogen produced in the shell burning.

In this paper, we investigate whether such low mass Pop III stars
ranging from 30 to 90$M_{\odot}$ can be progenitors of GRBs.  For this
purpose, we consider the jet propagation in the stellar envelope and
analytically calculate the evolution of the jet-cocoon structure.  In
BSG envelopes, the jet head velocity is larger than the cocoon
velocity all the way except for the very early time and we can expect
a successful jet breakout.  On the other hand, in RSG envelopes, the
cocoon edge reaches the stellar surface as early as or even earlier
than the jet head.  We confirm that Pop III RSGs have enough largely
extended envelopes for jets to be stalled on the way and cannot raise
GRBs as shown in \cite{2003MNRAS.345..575M}.  We also confirm that Pop
III BSGs are compact enough for the successful jet breakout and have
the possibility to raise GRBs as suggested in
\cite{2001ApJ...556L..37M} and \cite{2012ApJ...752...32W}.
%
It should be noted that the BSG models from Woosley and Heger used
above ignored the effect of the rotation on the stellar
evolution. \cite{2008A&A...489..685E} found that when the rotation is
included, Pop III stars within our noticed mass range end up as RSGs
not BSGs. But \cite{2008A&A...489..685E} considered the evolution of a
star with an extremely high rotation velocity as one half the critical
velocity. Recent cosmological simulations (\citealt{2010MNRAS.403..45}
and \citealt{2011ApJ...727..110C}), however, suggested that Pop III
stars are born in binary systems. In this case, the angular momentum
which the star forming gas clump initially has is divided into the
spin of each star and the orbital angular momentum so that these
stars may rotate less rapidly. Therefore, we think that such rapidly
rotating stars they considered are rare and the calculations based on
BSG models from Woosley and Heger make sense.

Using our model, we evaluate observational characters of Pop III
GRBs. We predict that although Pop III GRBs radiate as much energy as
the most energetic local long GRBs, Pop III GRBs are slightly less
luminous than local long GRBs due to their much longer burst duration.
Assuming that the $E_p-L_p$ (or $E_p-E_{\gamma,{\rm iso}}$)
correlation holds for Pop III GRBs, we predict that Pop III GRBs have
the much softer (or mildly softer) spectra than local long GRBs in the
observer frame.  \cite{2012ApJ...752...32W} predicted that the
gamma-ray transients from low metallicity BSGs have duration of
$10^{4-5}$ sec and the luminosity of $10^{48-49}$ erg ${\rm
  sec}^{-1}$, similar to Pop III GRBs considered here. Their
transients are fed by the mass accretion of the outer most layers of
stars and the accretion rate onto the BH ($\sim 10^{-4} M_{\odot}
\ {\rm sec}^{-1}$) is much smaller than that in the Pop III GRB case,
which imply typically
from $\sim 10^{-3} M_{\odot} \ {\rm sec}^{-1}$ to $\sim 10^{-2}
M_{\odot} \ {\rm sec}^{-1}$.  Moreover, they simply assumed a central
engine model with the roughly estimated conversion efficiency from the
mass accretion to the jet energy as $\sim 0.1$, while we consider a
central engine which is driven by the magnetic process implicitly
taking the disk accretion into account with the efficiency of $\eta
\sim 6.2 \times 10^{-4}$.\footnote{Note that
  \cite{2012ApJ...752...32W} considered the rotationally supported
  disk structure and the mass accretion from it so that the meaning of
  the conversion efficiency is different from ours.} Note that we
choose this value so as for Wolf-Rayet stars to reproduce the
energetics of local long GRBs.  Therefore, although the characters are
similar among them, we expect that gamma-ray transients considered in
\cite{2012ApJ...752...32W} are different events from GRBs considered
in this paper.

We also discuss the detectability of Pop III GRBs by future satellite
missions such as {\it Lobster} and {\it EXIST} in detail. If the
$E_p-E_{\gamma,{\rm iso}}$ correlation holds, we have the possibility
to detect Pop III GRBs at redshifts $z \sim 9$ as long duration X-ray
rich GRBs by {\it EXIST}.  On the other hand, if the $E_p-L_p$
correlation holds, we have the possibility to detect Pop III GRBs up
to $z \sim 19$ as long duration X-ray flashes by {\it Lobster}.

We briefly comment the expected observable GRB rate per year by {\it
  Lobster} using the results of \cite{2011AA...533A..32D}.  We
calculate the observed GRB rate per year $dN_{\rm GRB}^{\rm obs}/dz$
as
\begin{equation}
\frac{dN_{\rm GRB}^{\rm obs}}{dz} = \frac{\Omega_{\rm obs}}{4 \pi} \eta_{\rm beam} \frac{dN_{\rm GRB}}{dz},
\end{equation}
where $dN_{\rm GRB}/dz$, $\Omega_{\rm obs}$ and $\eta_{\rm beam}$
correspond to the intrinsic GRB rate (the number of on-axis and
off-axis GRBs) per year, the detector field of view and the beaming
factor of the burst. In Fig. 6 of \cite{2011AA...533A..32D}, they
showed $dN_{\rm GRB}/dz$ for an {\it optimistic} case and we use
their values. Here, we also adopt the values of $\eta_{\rm beam} \sim
0.01$ and $\Omega_{\rm obs} \sim 0.5$ sr for {\it Lobster}
\citep{2012IAUS..285...41G}. Optimistically speaking, we predict that
{\it Lobster} detects about 40, 4 and 0.4 Pop III GRBs per year at 
$z=9, 14$ and 19, respectively.

At last, we briefly discuss employed assumptions in this paper.
Firstly, we assume that all the stars considered in this paper
($30-90M_{\odot}$ Pop III stars) form black holes directly after the
stellar core collapse (see \S3.3). It should be noted, however, that
this is not always the case especially for less massive stars. Shortly
after the onset of the core collapse, a neutron star and a shock wave,
which propagates outward or is stalled, are considered to be formed at
first. Behind the shock wave, a fall back accretion of the shocked
envelope could be present and the continuous accretion onto the
neutron star eventually leads to a black hole formation. Although the
early activity of the central engine could be affected by the
accretion details, i.e., the direct accretion or fall-back accretion,
the conclusion of this paper is hardly changed. This is because the
mass accretion at the interested time in this paper is coming from the
massive envelope so that the central engine already collapsed to a
black hole at the corresponding time. In addition, since the energy
budget of the shock head is dominated by the envelope accretion, the
details of the early phase does not affect the shock evolution in the
late phase.  For a more massive ($\gtrsim 40M_{\odot}$) star, on the
other hand, the energy of the shock wave is too low to explode even
the portion of the envelope, so a black hole would be formed directly
and our assumption is fully justified in this case. Note that the mass
threshold between the direct or fall-back induced black hole formation
is still under the debate (see e.g. \citealt{1999ApJ...522..413F}) and
beyond the scope of this paper.
Secondly, we assume that the whole stellar envelope accretes onto the
BH (see \S3.1). In order to confirm the validity of this assumption,
we evaluate the binding energy of each layer of the stellar envelope
and compare it with the typical energy of a supernova outgoing shock
$\sim 10^{51}$ erg, which is injected around $\sim 10$ km from the
center. We find that the binding energy becomes larger than $10^{51}$
erg within $r \lesssim 5 \times 10^9$ cm. This means that the outgoing
shock should stall on the way and we expect little mass
ejection. Recently, \cite{2012MNRAS.423L..92Q} suggested that in the
late stage of the stellar evolution, the pre-supernova burning leads
to a significant mass ejection from the outer envelope. But they
considered only the case of a $40 M_{\odot}$ star with metallicity $Z
= 10^{-4}$ and commented that the amount of mass ejection depends on
the metallicity and rotation etc. Thus, the amount of the ejectable
mass is uncertain for the progenitors employed here so that we do not
consider this effect in this paper.

\section*{Acknowledgements}
We thank A. Heger for kindly providing us his stellar model data. We
thank D. Yonetoku and R. Yamazaki for fruitful discussion and
suggestions about GRB observations. We also thank K. Omukai for
fruitful discussions about Population III stars and the anonymous referee for fruitful comments. This work is
supported in part by the Grant-in-Aid from the Ministry of Education,
Culture, Sports, Science and Technology (MEXT) of Japan, No.23540305
(TN), No.24103006 (TN), No.23840023(YS) and by the Grant-in-Aid for
the global COE program {\it The Next Generation of Physics, Spun from
  Universality and Emergence} at Kyoto University.

\newpage

\begin{table}[!htb]
\caption{The comparison of typical long GRBs and Pop III GRBs with $40M_{\odot}$.}
\begin{center}
\begin{tabular}{c|cc}
\hline
                                                        &   long GRB         &  Pop III GRB ($z=9$)      \\ \hline
$E_{\gamma, {\rm iso}} $ [erg]        &  $10^{52-54}$    &  $10^{54}$                         \\ 
$L_{\rm p}$ [erg\ ${\rm sec}^{-1}$]                 &  $10^{52-53}$    &  $6 \times 10^{50}$                         \\ 
$T_{90}$ [sec]                                 &  $10^{1-3}$      &  $6 \times 10^{4}$                           \\
$E_p^{\rm obs}$ [keV]                      &  $10^{2-3}$      &  $5.5$  ($E_p-L_p$)             \\ 
                                                        &                            &  $120$ ($E_p-E_{\gamma,{\rm iso}}$)             \\
\hline
\end{tabular} 
\end{center}
\label{tab:popIIIGRB}
\end{table}

\begin{table}[!htb]
\caption{The observational characteristics of Pop III GRBs at $z=19$ for various progenitor masses.}
\begin{center}
\begin{tabular}{c|cccc}
\hline
mass [$M_{\odot}$]                                      & $30$ & $44$        & $60$& $70$ \\ \hline
$E_{\gamma, {\rm iso}}$ ($10^{54}$) [erg]    & $0.94$        & $1.1 $ & $1.5 $ & $1.6 $       \\ 
$L_{\rm p}$ ($10^{50}$) [erg\ ${\rm sec}^{-1}$]                  & $5.1$ & $6.4$ & $1.7$ & $1.9$                      \\ 
$T_{90}$ ($10^{5}$) [sec]                             &   $0.87$        & $1.3$ & $10$ & $11$   \\
$E_p^{\rm obs}$ [keV] ($E_p-L_p$)                    & $2.5$          & $2.8$ & $1.4$ & $1.5$                \\ 
$E_p^{\rm obs}$ [keV] ($E_p-E_{\gamma,{\rm iso}}$)   & $54$                & $60$ & $70$ & $75$              \\
\hline
\end{tabular} 
\end{center}
\label{tab:obsmass}
\end{table}

\end{document}